\documentclass[final,5p,times,twocolumn]{elsarticle}
\usepackage{hyperref}
\usepackage{amsmath}
\usepackage{amssymb}
\usepackage{graphicx}

\begin{document}

\begin{frontmatter}

\title{Heat generation due to spin transport in spin valves}

\author{Xiao-Xue Zhang}

\author{Pei-Song He}
\author{Bao-He Li}
\author{Yao-Hui Zhu\corref{mycorrespondingauthor}}
\cortext[mycorrespondingauthor]{Corresponding author}
\ead{yaohuizhu@gmail.com}
\address{Physics Department, Beijing Technology and Business University, Beijing 100048, China}

\begin{abstract}
Using a macroscopic approach, we studied theoretically the heat generation due to spin transport in a typical spin valve with nonmagnetic spacer layer of finite thickness. Our analysis shows that the spin-dependent heat generation can also be caused by another mechanism, the spin-conserving scattering in the presence of spin accumulation gradient, in addition to the well-known spin-flip scattering. The two mechanisms have equal contributions in semi-infinite layers, such as the ferromagnetic layers of the spin valve. However, in the nonmagnetic layer of a thickness much smaller than its spin-diffusion length, the spin-dependent heat generation is dominated by the spin-flip scattering in the antiparallel configuration, and by the spin-conserving scattering in the parallel configuration. We also proved that the spin-dependent heat generation cannot be interpreted as the Joule heating of the spin-coupled interface resistance in each individual layer. An effective resistance is proposed as an alternative so that the heat generation can still be described simply by applying Joule's law to an equivalent circuit.
\end{abstract}


\end{frontmatter}


\section{Introduction}

Heat generation is a serious issue even for spintronic devices.~\cite{Nobel2008,Bauer2012,Adachi2013} For example, large current is usually required for the operation of a spin-transfer-torque magnetic random-access memory.~\cite{Ralph2008} The reduction of the working current and the associated heating is still a challenging problem. Moreover, heating in spintronic devices leads to temperature gradient, which may inversely have remarkable influence on the spin transport via the spin-dependent Seebeck effect.~\cite{Seebeck2008}

Recent theoretical investigations have shown that there is still dissipation even if a pure spin current is present.~\cite{T:Boltz2011,Wegrowe2011,Slachter2011Nov,T:pump2014,Juarez16} Meanwhile, experimental studies have also demonstrated various spin-dependent heating effects.~\cite{Slachter2011,Krause2011,E:valve2012,E:resis2010,E:T-O2014,Peltier2012} It has been well established that heat generation in spintronic devices is dependent on the spin accumulation and spin current in comparison to that in conventional electronic devices. However, previous works did not pay much attention to the distinctions and characteristics of the various mechanisms causing the heat generation, especially the one due to the \emph{spin-conserving} scattering. This mechanism can also lead to spin-dependent heat generation when electrons flow to positions with lower chemical potential without spin flip. We will show that this mechanism can be identified if the equation of the heat generation is written in a more appropriate form, in which all terms are exactly the products of current-force pairs. Unfortunately, not all of the current-force pairs are correctly chosen in the previous papers. Furthermore, it is worthwhile to investigate in depth the relative magnitude of heat generation due to spin-conserving and spin-flip scattering, especially in a layer with finite thickness, such as the nonmagnetic (NM) spacer layer in a spin valve. The variation of heat production with the thickness of the NM layer is also an important problem.

From a macroscopic viewpoint, Ref.~\cite{T:Boltz2011} also showed that Joule heating of the spin-coupled interface resistance ($r_\mathrm{SI}$) is equal to the additional heat generation due to spin transport in a spin valve without a NM spacer layer. It is necessary to study whether this conclusion holds generally for spin valves with NM spacer layers of finite thickness. This question is closely related to the nonlocal character of the heat generation, which is similar to the situation of spin pumping discussed in Ref.~\cite{T:pump2014}. It is well-known that the NM layer in a spin valve has no contribution to the spin-coupled interface resistance because of the absence of the extra field.~\cite{Fert1993,Zhu14} However, the NM layer contributes to the additional heat generation due to spin transport. Thus the heat generation does not obey Joule's law in each individual layer, although it does throughout the whole structure. This character needs to be interpreted properly. In view of these open questions, we studied analytically the heat generation due to spin transport in a spin valve using a macroscopic approach based on the Boltzmann equation.~\cite{T:Boltz2011}

This paper is organized as follows. In Sec.~\ref{sec:Basic-theories-1}, we
derive the basic equations for the heat generation in magnetic multilayers and find the general relation between the spin-dependent heat generation and the spin-coupled interface resistance. Then the basic equations are applied to a spin valve in Sec.~ \ref{sec:Joule-heating-in-1}. We also compare the contribution from the spin-conserving and spin-flip scattering. In Sec.\ref{sec:Relation-with-the}, we discuss in depth the validity of the spin-coupled interface resistance and introduce the effective resistance as an alternative. Finally, our main results are summarized in Sec.~\ref{sec:Conclusions-1}.

\section{Basic equations\label{sec:Basic-theories-1}}

We are mainly concerned with the heat generation in a spin valve driven only by a constant current of density $J$, which flows in the $z$-direction and perpendicular to the layer plane.~\cite{Fert1993} In steady state, the time rate of heat generation can be calculated by using a macroscopic equation like Eq.~(6) in Ref.~\cite{T:Boltz2011}
\begin{equation}\label{eq:entropy2}
\sigma_\mathrm{heat}=\frac{J_{+}}{e}\frac{\partial\bar{\mu}_{+}}{\partial z}+\frac{J_{-}}{e}\frac{\partial\bar{\mu}_{-}}{\partial z}+\frac{4(\Delta\mu)^2}{e^2}G_\mathrm{mix}
\end{equation}
where we use $\sigma_\mathrm{heat}$ to denote the heat-generation rate following Ref.~\cite{Kondepudi}. This kind of equation can be derived by using the Boltzmann equation~\cite{T:Boltz2011} as well as nonequilibrium thermodynamics~\cite{Kondepudi} in the linear regime. In Eq.~\eqref{eq:entropy2}, $J_{+}$ ($J_{-}$) and $\bar\mu_{+}$ ($\bar\mu_{-}$) are the current density and the electro-chemical potential in spin-up (down) channel, respectively. The electron-number currents are given by $-J_\pm/e$, where $-e$ is the charge of an electron. Moreover, $\Delta\mu=(\bar\mu_{+}-\bar\mu_{-})/2$ describes the spin accumulation, and $G_\mathrm{mix}$ defined by Eq.~(5) of Ref.~\cite{T:Boltz2011} stands for the associated spin-flip rate.

The first two terms of Eq.~(\ref{eq:entropy2}) can be regarded as the decrease in the (electrochemical) potential energy current or the heat generation, in each spin channel.~\cite{Callen} The heat-generation rates of the two channels are unequal in ferromagnetic (FM) layers and at spin-selective interfaces. If the two channels cannot exchange heat effectively with each other or other heat reservoir, they may have different temperatures~\cite{Hatami07}. However, this is beyond the scope of the present work and we neglect this effect by assuming that the two spin channels can exchange energy effectively (the thermalized regime~\cite{Hatami07}). Then it is more meaningful to write Eq.~(\ref{eq:entropy2}) in terms of $J=J_{+}+J_{-}$ and $J_\mathrm{spin}=J_{+}-J_{-}$ like Eq.~(12c) of Ref.~\cite{T:Boltz2011}
\begin{equation}\label{eq:heat2}
\sigma_\mathrm{heat}=\frac{J}{e}\frac{\partial\bar{\mu}}{\partial z}
+\frac{J_\mathrm{spin}}{e}\frac{\partial\Delta{\mu}}{\partial z}
+\frac{\partial{J}_\mathrm{spin}}{\partial{z}}\frac{\Delta\mu}{e}
\end{equation}
where $\bar\mu=(\bar{\mu}_{+}+\bar{\mu}_{-})/2$ is the average electrochemical potential. Although each term of $\sigma_\mathrm{heat}$ has been written as the product of generalized force and current, which is consistent with the requirement of nonequilibrium thermodynamics,~\cite{Kondepudi} the current-force pairs in the first two terms of Eq.~\eqref{eq:heat2} are not exact pairs and need to be rewritten in a more appropriate form.

In FM layers, the total current density $J$ does not depend on position $z$ and thus it is not exactly the current corresponding to the force ${\partial\bar{\mu}}/{\partial z}$, which has exponential terms of $z$. This becomes obvious if we sum the `$\pm$' components of Eq.~\eqref{eq:conti:Ohm's law2}
\begin{equation}
F=\frac{1}{e}\frac{\partial\bar{\mu}}{\partial{z}}=E_0^\mathrm{F}
\pm\frac{\beta}{e}\frac{\partial\Delta\mu}{\partial{z}}\label{eq:meanfield}
\end{equation}
in an FM layer with `up' (`down') magnetization and bulk spin asymmetry coefficient $\beta$. Here, $\beta$ and the spin-dependent resistivity $\rho_{\uparrow(\downarrow)}$ [or conductivity $\sigma_{\uparrow(\downarrow)}$] satisfy the relation $\rho_{\uparrow(\downarrow)}=1/\sigma_{\uparrow(\downarrow)}=2\rho_\mathrm{F}^\ast[1-(+)\beta]$.~\cite{Fert1993} Note that the subscript ``$\uparrow$'' (``$\downarrow$'') denotes the majority (minority) spin direction. In Eq.~\eqref{eq:meanfield}, the effective field $F(z)$ of the FM layers has been separated into the bulk term $E_0^\mathrm{F}=J(1-\beta^2)\rho_\mathrm{F}^\ast$ and the exponential term. Meanwhile, the spin current $J_\mathrm{spin}$ has a bulk term and thus it is not exactly the current corresponding to the force ${\partial\Delta{\mu}}/{\partial z}$, which has only exponential terms. One can see this easily by subtracting the `$\pm$' components of Eq.~\eqref{eq:conti:Ohm's law2}
\begin{equation}\label{spincurrent}
J_\mathrm{spin}=\mp\beta{J}+\frac{1}{e\rho_\mathrm{F}^\ast}
\frac{\partial\Delta\mu}{\partial{z}}
\end{equation}
where the FM layer also has `up' (`down') magnetization. The total spin current in Eq.~\eqref{spincurrent}, $J_\mathrm{spin}$, has been written as the sum of a bulk term and an exponential one
\begin{align}
J_\mathrm{spin}^\mathrm{bulk}&=\mp\beta{J}\\
J_\mathrm{spin}^\mathrm{exp}&=\frac{1}{e\rho_\mathrm{F}^\ast}
\frac{\partial\Delta\mu}{\partial{z}}\label{spinexp}
\end{align}
Therefore, it is necessary to transform the first two terms of $\sigma_\mathrm{heat}$ in Eq.~(\ref{eq:heat2}) into exact current-force pairs. Fortunately, this desired form can be achieved by simply substituting Eqs.~(\ref{eq:meanfield}) and (\ref{spincurrent}) into Eq.~(\ref{eq:heat2})
\begin{equation}\label{eq:heat3}
\sigma_\mathrm{heat}^\mathrm{bulk}+\sigma_\mathrm{heat}^\mathrm{sc}
=JE_0^\mathrm{F}+\frac{J_\mathrm{spin}^\mathrm{exp}}{e}\frac{\partial\Delta\mu}{\partial{z}}
\end{equation}
where we have introduced
\begin{equation}\label{heatsc}
\sigma_\mathrm{heat}^\mathrm{sc}=
\frac{J_\mathrm{spin}^\mathrm{exp}}{e}\frac{\partial\Delta\mu}{\partial{z}}
\end{equation}
The first term, $\sigma_\mathrm{heat}^\mathrm{bulk}=JE_0^\mathrm{F}$, in Eq.~(\ref{eq:heat3}) has become the product of a current-force pair and it means the bulk Joule heating. The second term, $\sigma_\mathrm{heat}^\mathrm{sc}$, has also been written as the product of a current and its corresponding force, and it can be regarded as the heat generation due to the spin-conserving (sc) scattering. The reason is that, the gradient of spin accumulation drives a spin from a position to another one with lower chemical potential via the spin-conserving scattering, and the excess chemical potential of the spin leads to heat generation. This can also be understood by writing this term as the sum of the contributions from the two spin channels (see~\ref{spindiffusion}). The heat generation due to spin-conserving scattering $\sigma_\mathrm{heat}^\mathrm{sc}$ did not attract much attention in the previous studies, such as Refs.~\cite{T:Boltz2011,T:pump2014}. This mechanism is especially important for the heat generation due to the interface resistance because the spin-flip scattering is usually neglected at interfaces. We will look into $\sigma_\mathrm{heat}^\mathrm{sc}$ in Sec.~\ref{sec:Joule-heating-in-1}, and compare it with $\sigma_\mathrm{heat}^\mathrm{sf}$ (see below) in typical spin valves.

Next we will rewrite the last term of $\sigma_\mathrm{heat}$ of Eq.~(\ref{eq:heat2}) in a form with more obvious physical interpretation. Subtracting the `$\pm$' components of Eq.~\eqref{eq:conti:spin-flip2} in two different ways, one can derive
\begin{equation}\label{jspinderivative}
\frac{\partial{J}_\mathrm{spin}}{\partial{z}}
=\frac{4eN_s\Delta\mu}{\tau_\mathrm{sf}^\mathrm{F(N)}}
=\frac{\Delta\mu}{er_\mathrm{F(N)}l_\mathrm{sf}^\mathrm{F(N)}}
\end{equation}
where $r_\mathrm{F}=\rho_\mathrm{F}^\ast{l}_\mathrm{sf}^\mathrm{F}$ and $r_\mathrm{N}=\rho_\mathrm{N}^\ast{l}_\mathrm{sf}^\mathrm{N}$. In Eq.~\eqref{jspinderivative}, $N_s$ is the density of states for spin $s$ and $\tau_\mathrm{sf}^\mathrm{F}$ ($\tau_\mathrm{sf}^\mathrm{N}$) the spin-flip relaxation time of the FM (NM) layer. Then using Eq.~\eqref{jspinderivative}, we can rewrite the last term of Eq.~(\ref{eq:heat2}) as
\begin{equation}\label{heatsf}
\sigma_\mathrm{heat}^\mathrm{sf}=\alpha_\mathrm{spin}\mu_\mathrm{m}
\end{equation}
where $\alpha_\mathrm{spin}=\mu_\mathrm{m}{N}_s/{\tau_\mathrm{sf}}$ and $\mu_\mathrm{m}=2\Delta\mu$. Here $\alpha_\mathrm{spin}$
describes the damping rate of the spin accumulation, that is, the number of electrons flipping from spin-up to spin-down states per unit time per unit volume, which is called the spin flux $\dot{\psi}$ in Ref.~\cite{Wegrowe2011}. Thus $\alpha_\mathrm{spin}$ and $\mu_\mathrm{m}$ can also be regarded as a generalized current-force pair, and $\sigma_\mathrm{heat}^\mathrm{sf}$ stands for the heat generation due to the spin-flip (sf) scattering.

Collecting all the terms, we can then write the rate of heat generation in FM (NM) layer in the form of Joule's law
\begin{equation}\label{eq:heatf}
\begin{split}
\sigma_\mathrm{heat}&=\sigma_\mathrm{heat}^\mathrm{bulk}
+\sigma_\mathrm{heat}^\mathrm{sc}
+\sigma_\mathrm{heat}^\mathrm{sf}\\
&=(1-\beta^2)\rho_\mathrm{F(N)}^\ast{J}^2
+\rho_\mathrm{F(N)}^\ast\left(J_\mathrm{spin}^\mathrm{exp}\right)^2
+\rho_\mathrm{F(N)}^\ast\left[\frac{\Delta\mu}{er_\mathrm{F(N)}}\right]^2
\end{split}
\end{equation}
where we have rewritten $\sigma_\mathrm{heat}^\mathrm{sc}$ and $\sigma_\mathrm{heat}^\mathrm{sf}$ using Eqs.~\eqref{spinexp} and \eqref{jspinderivative}. The first term $\sigma_\mathrm{heat}^\mathrm{bulk}$ is due to the nominal resistance and exists no matter whether the current is spin-polarized or not. However, $\sigma_\mathrm{heat}^\mathrm{sc}$ and $\sigma_\mathrm{heat}^\mathrm{sf}$ rely on the \emph{gradients} of the spin accumulation and spin current according to Eqs.~\eqref{spinexp} and \eqref{jspinderivative}, respectively.

Finally, we will discuss the general relation between the spin-dependent heat generation and the `Joule heating' of the spin-coupled interface resistance.~\cite{T:Boltz2011,Fert1993} To this purpose, we write Eq.~(\ref{eq:heat2}) as
\begin{equation}\label{eq:heat4}
\sigma_\mathrm{heat}=JF
+\frac{1}{e}\frac{\partial}{\partial{z}}\left(J_\mathrm{spin}\Delta{\mu}\right)
\end{equation}
by using $F=(1/e)\partial\bar\mu/\partial{z}$. Without loss of generality, we will consider a segment of a multilayer from $z_\mathrm{L}$ to $z_\mathrm{R}$, which includes an interface at $z_\mathrm{C}$ ($z_\mathrm{L}<z_\mathrm{C}<z_\mathrm{R}$). Integrating Eq.~\eqref{eq:heat4} from $z_\mathrm{L}$ to $z_\mathrm{R}$, we can write the total heat generation in this segment as
\begin{equation}\label{totalhg}
\Sigma_\mathrm{heat}=\Sigma_\mathrm{heat}^\mathrm{L}
+\Sigma_\mathrm{heat}^\mathrm{C}+\Sigma_\mathrm{heat}^\mathrm{R}
\end{equation}
where
\begin{align}
\Sigma_\mathrm{heat}^\mathrm{L}&=J\Delta{V}_0^\mathrm{L}+J\Delta{V}_\mathrm{I}^\mathrm{L}
+\frac{1}{e}\left(J_\mathrm{spin}\Delta{\mu}\right)\bigg|_{z_\mathrm{L}}^{z_\mathrm{C}^-}\label{totalhl}\\
\Sigma_\mathrm{heat}^\mathrm{C}&=\frac{J}{e}\left[\bar\mu(z_\mathrm{C}^+)-\bar\mu(z_\mathrm{C}^-)\right]
+\frac{1}{e}\left(J_\mathrm{spin}\Delta{\mu}\right)\bigg|_{z_\mathrm{C}^-}^{z_\mathrm{C}^+}\label{totalhir}\\
\Sigma_\mathrm{heat}^\mathrm{R}&=J\Delta{V}_0^\mathrm{R}+J\Delta{V}_\mathrm{I}^\mathrm{R}
+\frac{1}{e}\left(J_\mathrm{spin}\Delta{\mu}\right)\bigg|_{z_\mathrm{C}^+}^{z_\mathrm{R}}\label{totalhr}
\end{align}
are defined as the integral heat-generation rate for the left layer, the interface, and the right layer, respectively. In Eqs.~\eqref{totalhl}, $\Delta{V}_0^\mathrm{L}$ and $\Delta{V}_\mathrm{I}^\mathrm{L}$ are defined as
\begin{align}
\Delta{V}_0^\mathrm{L}&=\int_{z_\mathrm{L}}^{z_\mathrm{C}}E_0^\mathrm{L}dz
=J\left(1-\beta^2\right)\rho_\mathrm{L}^\ast\left(z_\mathrm{C}-z_\mathrm{L}\right)\\
\Delta{V}_\mathrm{I}^\mathrm{L}&=Jr_\mathrm{SI}^\mathrm{L}
=\int_{z_\mathrm{L}}^{z_\mathrm{C}}\left(F-E_0^\mathrm{L}\right)dz
=\pm\beta\frac{\Delta\mu(z_\mathrm{C}^-)-\Delta\mu(z_\mathrm{L})}{e}\label{rsi}
\end{align}
where we have used Eq.~\eqref{eq:meanfield}. Similarly, $\Delta{V}_0^\mathrm{R}$ and $\Delta{V}_\mathrm{I}^\mathrm{R}$ in \eqref{totalhr} are defined in the regime $z_\mathrm{C}<z<z_\mathrm{R}$. In Eq.~(\ref{rsi}), we have defined $r_\mathrm{SI}^\mathrm{L}$ as the spin-coupled interface resistance of the left layer by extending the definition in Ref.~\cite{Fert1993}.

The interface has been treated as a layer with an infinitesimal thickness in Eq.~\eqref{totalhir}. This approach is equivalent to the method used in Ref.~\cite{T:Boltz2011}, which is proven in \ref{appir}. Using Eqs.~\eqref{bcif1}, \eqref{deltamu}, and \eqref{deltamubar}, we can rewrite Eq.~\eqref{totalhir} as
\begin{equation}
\Sigma_\mathrm{heat}^\mathrm{C}=J\Delta{V}_0^\mathrm{C}+J\Delta{V}_\mathrm{I}^\mathrm{C}
+J_\mathrm{spin}(z_\mathrm{C})\frac{\Delta\mu(z_\mathrm{C}^+)-\Delta\mu(z_\mathrm{C}^-)}{e}
\label{totalhir2}
\end{equation}
where $\Delta{V}_0^\mathrm{C}$ and $\Delta{V}_\mathrm{I}^\mathrm{C}$ are defined as
\begin{align}
\Delta{V}_0^\mathrm{C}&=J(1-\gamma^2)r_\mathrm{b}^\ast\label{rifnominal}\\
\Delta{V}_\mathrm{I}^\mathrm{C}&=Jr_\mathrm{SI}^\mathrm{C}
=\pm\gamma\frac{\Delta\mu(z_\mathrm{C}^+)-\Delta\mu(z_\mathrm{C}^-)}{e}\label{rsiif}
\end{align}
In Eq.~\eqref{rsiif}, $r_\mathrm{SI}^\mathrm{C}$ is defined as the spin-coupled interface resistance of the interface. The sign `$+$' (`$-$') in Eq.~\eqref{rsiif} corresponds to the configuration where the spin-up channel is the minority (majority) one. Substituting Eq.~\eqref{totalhir2} into Eq.~\eqref{totalhg}, we have
\begin{equation}\label{heattotal}
\begin{split}
\Sigma_\mathrm{heat}&=J\left(\Delta{V}_0^\mathrm{L}+\Delta{V}_0^\mathrm{C}
+\Delta{V}_0^\mathrm{R}\right)
+J\left(\Delta{V}_\mathrm{I}^\mathrm{L}+\Delta{V}_\mathrm{I}^\mathrm{C}
+\Delta{V}_\mathrm{I}^\mathrm{R}\right)\\
&+\left[J_\mathrm{spin}(z_\mathrm{R})\Delta{\mu}(z_\mathrm{R})
-J_\mathrm{spin}(z_\mathrm{L})\Delta{\mu}(z_\mathrm{L})\right]/e
\end{split}
\end{equation}
where the first term on the right-hand side is the nominal Joule heating and the second the `Joule heating' of the spin-coupled interface resistance. The last term of Eq.~\eqref{heattotal} disappears once $J_\mathrm{spin}\Delta\mu$ has the same value at $z_\mathrm{L}$ and $z_\mathrm{R}$.

On the other hand, by using Eqs.~\eqref{eq:meanfield} and \eqref{spincurrent}, one can also rewrite Eq.~\eqref{eq:heat2} as
\begin{equation}\label{eq:heat5}
\sigma_\mathrm{heat}=JE_0^\mathrm{F(N)}+\frac{1}{e}\frac{\partial}{\partial{z}}
\left(J_\mathrm{spin}^\mathrm{exp}\Delta\mu\right)
\end{equation}
where the last term is the sum of $\sigma_\mathrm{heat}^\mathrm{sc}$ and $\sigma_\mathrm{heat}^\mathrm{sf}$. Then the total heat-generation rate in the segment $z_\mathrm{L}<z<z_\mathrm{R}$ can also be written as
\begin{equation}\label{heattotal2}
\Sigma_\mathrm{heat}=J\left(\Delta{V}_0^\mathrm{L}+\Delta{V}_0^\mathrm{C}
+\Delta{V}_0^\mathrm{R}\right)+\Sigma_\mathrm{heat}^\mathrm{spin}
\end{equation}
where
\begin{equation}\label{heatspin}
\Sigma_\mathrm{heat}^\mathrm{spin}=
\frac{1}{e}\left(J_\mathrm{spin}^\mathrm{exp}\Delta{\mu}\right)\bigg|_{z_\mathrm{L}}^{z_\mathrm{C}^-}
+\Sigma_\mathrm{heat}^\mathrm{spin,C}
+\frac{1}{e}\left(J_\mathrm{spin}^\mathrm{exp}\Delta{\mu}\right)\bigg|_{z_\mathrm{C}^+}^{z_\mathrm{R}}
\end{equation}
stands for the spin-dependent part of the total heat generation. The contribution of the interface can be written as
\begin{equation}
\Sigma_\mathrm{heat}^\mathrm{spin,C}=
\frac{1}{e}J_\mathrm{spin}^\mathrm{sa}(z_\mathrm{C})
\left[\Delta{\mu}(z_\mathrm{C}^+)-\Delta{\mu}(z_\mathrm{C}^-)\right]
\end{equation}
where $J_\mathrm{spin}^\mathrm{sa}(z_\mathrm{C})$ is defined by
\begin{equation}\label{spincurrentc}
J_\mathrm{spin}(z_\mathrm{C})
=\mp\gamma{J}+J_\mathrm{spin}^\mathrm{sa}(z_\mathrm{C})
\end{equation}
One can see that Eq.~\eqref{spincurrentc} is similar to Eq.~\eqref{spincurrent}. It is also easy to verify $\Sigma_\mathrm{heat}^\mathrm{C}=J\Delta{V}_0^\mathrm{C}+\Sigma_\mathrm{heat}^\mathrm{spin,C}$. Using Eq.~\eqref{deltamu}, we can rewrite $\Sigma_\mathrm{heat}^\mathrm{spin,C}$ in the form of Joule's law
\begin{equation}\label{heatcspin}
\Sigma_\mathrm{heat}^\mathrm{spin,C}=r_\mathrm{b}^\ast
\left[J_\mathrm{spin}^\mathrm{sa}(z_\mathrm{C})\right]^2
\end{equation}
which will be discussed further in Sec.~\ref{sub:Bulk-spin-dependent-scattering-1}.

Comparing Eqs.~\eqref{heattotal} and \eqref{heattotal2}, one can see that $\Sigma_\mathrm{heat}^\mathrm{spin}$ is equal to the `Joule heating' of the spin-coupled interface resistance
\begin{equation}\label{heatspin2}
\Sigma_\mathrm{heat}^\mathrm{spin}=J\left[\Delta{V}_\mathrm{I}^\mathrm{L}+\Delta{V}_\mathrm{I}^\mathrm{C}
+\Delta{V}_\mathrm{I}^\mathrm{R}\right]
\end{equation}
if the last term of Eq.~\eqref{heattotal} vanishes, that is,
\begin{equation}\label{requirement}
J_\mathrm{spin}(z_\mathrm{L})\Delta{\mu}(z_\mathrm{L})
=J_\mathrm{spin}(z_\mathrm{R})\Delta{\mu}(z_\mathrm{R})
\end{equation}
In fact, this requirement can be satisfied easily because a magnetic multilayer usually has several positions , where $J_\mathrm{spin}\Delta\mu$ is zero. We will discuss $r_\mathrm{SI}$ in more detail in Sec.~\ref{sec:Relation-with-the}.

\section{Heat generation in spin valves\label{sec:Joule-heating-in-1}}

In this section, we will apply the basic equations derived in Sec.~\ref{sec:Basic-theories-1} to spin valves with finite NM layer. The insertion of the NM layer enables us to study more realistically the difference in heat generation between parallel (P) and antiparallel (AP) alignments of the two FMs as well as the influence of the NM-layer thickness on the heat generation. To be specific, we place the origin of the $z$-axis at the center of the NM layer. The left and right FM/NM interfaces are located at $z=-d$ and $z=d$, respectively. The two semi-infinite FM layers are made of the same material with collinear magnetization. The various quantities in Eq.~\eqref{eq:heatf} can be derived according to Valet-Fert theory~\cite{Fert1993} and the detailed results are listed in~\ref{sec:Appendix:-General-Expression}. Substituting these results into Eq.~\eqref{eq:heatf}, one can get the time rate of heat generation in the left (`$+$') and right (`$-$') FM layers
\begin{equation}\label{heatfm1}
\sigma_\mathrm{heat}^\mathrm{P(AP)}=\left(1-\beta^{2}\right)\rho_\mathrm{F}^\ast{J}^{2}
+\Sigma_\mathrm{heat}^\mathrm{F,P(AP)}
\frac{2}{l_\mathrm{sf}^\mathrm{F}}\exp
\left[\pm\frac{2\left(z\pm{d}\right)}{l_\mathrm{sf}^\mathrm{F}}\right]
\end{equation}
where
\begin{equation}\label{totalh1}
\Sigma_\mathrm{heat}^\mathrm{F,P(AP)}=\int_{-\infty}^{-d}
\left[\sigma_\mathrm{heat}^\mathrm{sc,P(AP)}+\sigma_\mathrm{heat}^\mathrm{sf,P(AP)}\right]dz
=r_\mathrm{F}\left[\alpha_\mathrm{F}^\mathrm{P(AP)}J\right]^2
\end{equation}
is the spin-dependent part of the integral heat generation in the left FM layer. The right FM layer has the same contribution because both $\sigma_\mathrm{heat}^\mathrm{sc,P(AP)}$ and $\sigma_\mathrm{heat}^\mathrm{sf,P(AP)}$ are even functions about the origin. The dimensionless parameter $\alpha_\mathrm{F}^\mathrm{P(AP)}$ in Eq.~\eqref{totalh1} will be determined by boundary conditions in Sec.~\ref{sub:Bulk-spin-dependent-scattering-1}.

The heat generation in the NM layer is more complicated and has to be treated term by term. The bulk term $\sigma_\mathrm{heat}^\mathrm{bulk}$ has the same value $\sigma_\mathrm{heat}^\mathrm{bulk}=\rho_\mathrm{N}^\ast{J}^{2}$ for either AP or P alignment, and its integral over the NM layer ($-d<z<d$) can be regarded as the Joule heating of the NM-layer nominal resistance $2d\rho_{N}^{*}$. For the AP alignment, the two spin-dependent terms can be written as
\begin{align}
\sigma_\mathrm{heat}^\mathrm{sc,AP}&=
\frac{2\Sigma_\mathrm{heat}^\mathrm{N,AP}}{l_\mathrm{sf}^\mathrm{N}\sinh(2\xi)}
\sinh^2\left(\frac{z}{l_\mathrm{sf}^\mathrm{N}}\right)\label{heatnmapsc}\\
\sigma_\mathrm{heat}^\mathrm{sf,AP}&=
\frac{2\Sigma_\mathrm{heat}^\mathrm{N,AP}}{l_\mathrm{sf}^\mathrm{N}\sinh(2\xi)}
\cosh^2\left(\frac{z}{l_\mathrm{sf}^\mathrm{N}}\right)\label{heatnmapsf}
\end{align}
where
\begin{equation}\label{totalhnmap}
\Sigma_\mathrm{heat}^\mathrm{N,AP}=\int_{-d}^{0}
\left(\sigma_\mathrm{heat}^\mathrm{sc,AP}+\sigma_\mathrm{heat}^\mathrm{sf,AP}\right)dz
=r_\mathrm{N}^\mathrm{AP}\left(\alpha_\mathrm{N}^\mathrm{AP}J\right)^2
\end{equation}
is the spin-dependent part of the integral heat generation in the left half of the NM layer. The contribution from the right half is equal to that from the left half since both $\sigma_\mathrm{heat}^\mathrm{sc,AP}$ and $\sigma_\mathrm{heat}^\mathrm{sf,AP}$ are even functions. In Eq.~\eqref{totalhnmap}, $r_\mathrm{N}^\mathrm{AP}$ is defined as $r_\mathrm{N}^\mathrm{AP}=r_\mathrm{N}\coth\xi$, where we have $\xi=d/l_\mathrm{sf}^\mathrm{N}$. On the other hand, for the P alignment, we have
\begin{align}
\sigma_\mathrm{heat}^\mathrm{sc,P}&=
\frac{2\Sigma_\mathrm{heat}^\mathrm{N,P}}{l_\mathrm{sf}^\mathrm{N}
\sinh(2\xi)}\cosh^2\left(\frac{z}{l_\mathrm{sf}^\mathrm{N}}\right)\label{heatnmpsc}\\
\sigma_\mathrm{heat}^\mathrm{sf,P}&=
\frac{2\Sigma_\mathrm{heat}^\mathrm{N,P}}{l_\mathrm{sf}^\mathrm{N}\sinh(2\xi)}
\sinh^2\left(\frac{z}{l_\mathrm{sf}^\mathrm{N}}\right)\label{heatnmpsf}
\end{align}
where
\begin{equation}\label{totalhnmp}
\Sigma_\mathrm{heat}^\mathrm{N,P}=\int_{-d}^{0}
\left(\sigma_\mathrm{heat}^\mathrm{sc,P}+\sigma_\mathrm{heat}^\mathrm{sf,P}\right)dz
=r_\mathrm{N}^\mathrm{P}\left(\alpha_\mathrm{N}^\mathrm{P}J\right)^2
\end{equation}
is the spin-dependent part of the integral heat generation in the left half of the NM layer. Similarly, the right half has the same contribution due to symmetry. In Eq.~\eqref{totalhnmp}, $r_\mathrm{N}^\mathrm{P}$ is defined as $r_\mathrm{N}^\mathrm{P}=r_\mathrm{N}\tanh\xi$ and the dimensionless parameter $\alpha_\mathrm{N}^\mathrm{P(AP)}$ will be determined by boundary conditions in Sec.~\ref{sub:Bulk-spin-dependent-scattering-1}.

\subsection{Comparison of spin-conserving and spin-flip scattering \label{sec:The-contributions-of}}

Now we are ready to compare the magnitude of $\sigma_\mathrm{heat}^\mathrm{sc}$ and $\sigma_\mathrm{heat}^\mathrm{sf}$ in two typical situations: a semi-infinite layer and a finite layer. The spin valve contains both types of layers, namely the semi-infinite FM layers and the NM layer with a finite thickness. In the FM layers, the spin accumulation decays exponentially on the scale of the spin-diffusion length $l_\mathrm{sf}$ as shown in Eqs.~\eqref{deltamufm1} and \eqref{deltamufm2}. Without loss of generality, we will consider only the left FM layer ($z<-d$). Using Eq.~\eqref{eq:heatf}, one can easily verify
\begin{equation}
\sigma_\mathrm{heat}^\mathrm{sc,P(AP)}=\sigma_\mathrm{heat}^\mathrm{sf,P(AP)}
=\Sigma_\mathrm{heat}^\mathrm{F,P(AP)}
\frac{1}{l_\mathrm{sf}^\mathrm{F}}\exp
\left[\frac{2(z+d)}{l_\mathrm{sf}^\mathrm{F}}\right]
\end{equation}
which shows that the spin-conserving scattering has the same contribution to the heat generation at any position as the spin-flip scattering in semi-infinite layers. Thus their integrals from $-\infty$ to $-d$ are also equal to each other [see Eq.~\eqref{totalh1}].

\begin{figure}
\includegraphics[width=0.48\textwidth]{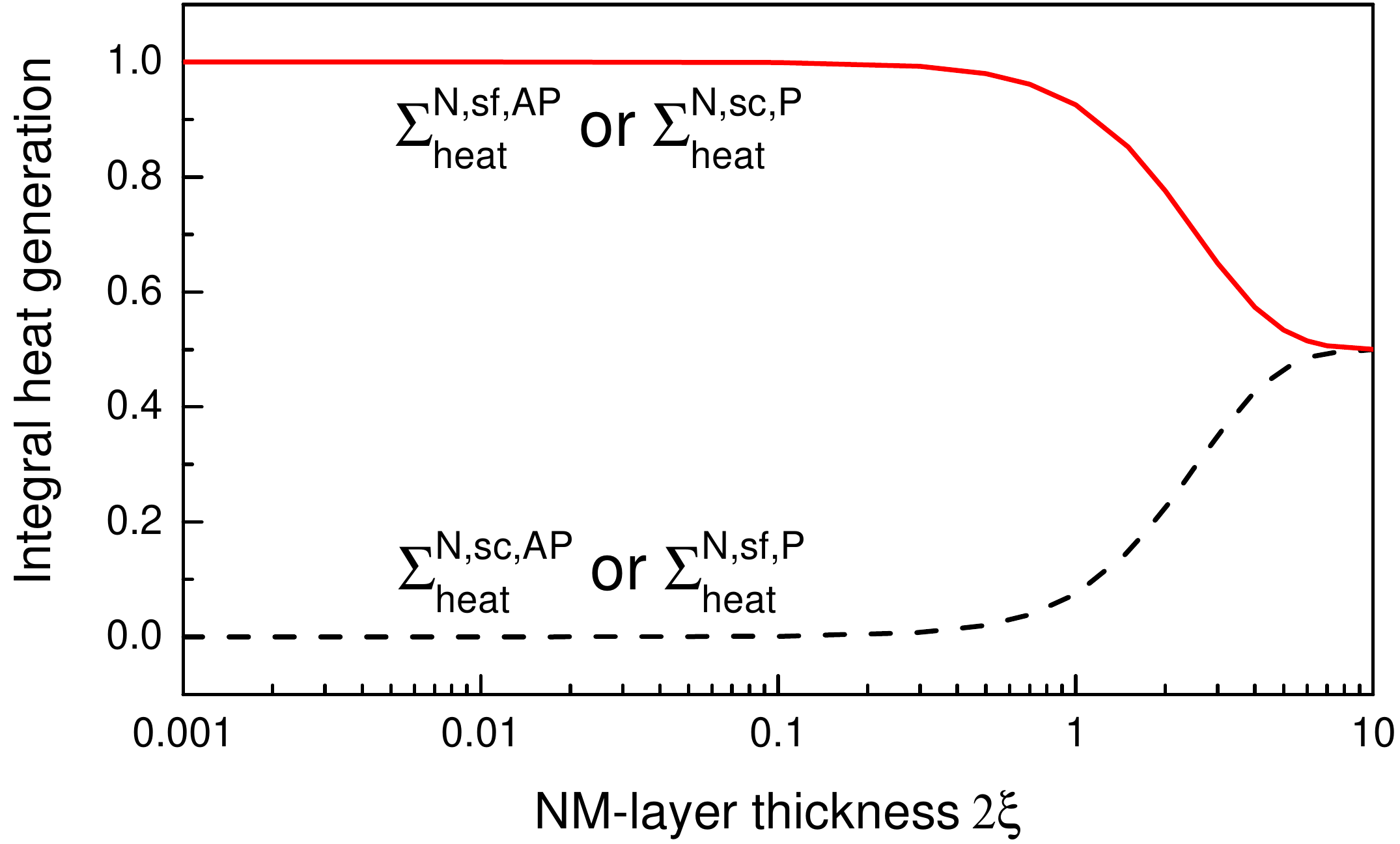}
\caption{\label{fig-relative}Integral heat generation rate (in unit of $\Sigma_\mathrm{heat}^\mathrm{N,AP}$ and $\Sigma_\mathrm{heat}^\mathrm{N,P}$ for AP and P alignments, respectively) in half of the NM layer as a function of the NM-layer thickness (in unit of $l_\mathrm{sf}^\mathrm{N}$). The red-solid curve corresponds to the integral heat generation due to the spin-flip (spin-conserving) scattering in the AP (P) configuration, and the black-dashed curve stands for that due to the spin-conserving (spin-flip) scattering in the AP (P) configuration.}
\end{figure}

As for the finite NM layer, it is more meaningful to compare the integral heat generation due to the spin-conserving and spin-flip scattering, denoted by $\Sigma_\mathrm{heat}^\mathrm{N,sc}$ and $\Sigma_\mathrm{heat}^\mathrm{N,sf}$, respectively. In the AP configuration, we have
\begin{align}
\Sigma_\mathrm{heat}^\mathrm{N,sc,AP}=\int_{-d}^{0}\sigma_\mathrm{heat}^\mathrm{sc,AP}dz
=\frac{1-\eta}{2}\Sigma_\mathrm{heat}^\mathrm{N,AP}\label{sigmanapsc}\\
\Sigma_\mathrm{heat}^\mathrm{N,sf,AP}=\int_{-d}^{0}\sigma_\mathrm{heat}^\mathrm{sf,AP}dz
=\frac{1+\eta}{2}\Sigma_\mathrm{heat}^\mathrm{N,AP}\label{sigmanapsf}
\end{align}
where the dimensionless parameter $\eta=2\xi/\sinh(2\xi)$ describes the asymmetry between $\Sigma_\mathrm{heat}^\mathrm{N,sc,AP}$ and $\Sigma_\mathrm{heat}^\mathrm{N,sf,AP}$. This becomes more obvious if one looks at their relative difference $(\Sigma_\mathrm{heat}^\mathrm{N,sf,AP}-\Sigma_\mathrm{heat}^\mathrm{N,sc,AP})
/\Sigma_\mathrm{heat}^\mathrm{N,AP}=\eta$,
which decreases from $1$ to $0$ as $\xi$ varies from $0$ to $\infty$. Figure~\ref{fig-relative} shows their variation with the NM layer thickness $2\xi$ (in unit of $l_\mathrm{sf}^\mathrm{N}$). In the regime of $2\xi=2d/l_\mathrm{sf}^\mathrm{N}\ll{1}$, which is practical for experiments, $\Sigma_\mathrm{heat}^\mathrm{N,sc,AP}/\Sigma_\mathrm{heat}^\mathrm{N,AP}$ and $\Sigma_\mathrm{heat}^\mathrm{N,sc,AP}/\Sigma_\mathrm{heat}^\mathrm{N,AP}$ approach $0$ and $1$, respectively. Therefore, the spin-dependent heat generation of the NM layer is dominated by the spin-flip scattering in the AP alignment. This feature can be interpreted as follows. We have $z/l_\mathrm{sf}^\mathrm{N}\ll{1}$ in the NM layer ($-d<z<d$) if $2d/l_\mathrm{sf}^\mathrm{N}\ll{1}$. This allows us to expand the spin accumulation $\Delta\mu$ [see Eq.~\eqref{deltamunm}] in terms of $z/l_\mathrm{sf}^\mathrm{N}$ and keep up to the first-order term. The result is a term independent of position because the first-order term also vanishes. Then $J_\mathrm{spin}^\mathrm{exp}$ approaches zero because it is proportional to the gradient of $\Delta\mu$ according to Eq.~\eqref{spincurrent}. Therefore, $\sigma_\mathrm{heat}^\mathrm{sc,AP}$ approaches zero in this regime according to Eq.~\eqref{heatsc}.

In the P configuration, the integral heat generation, $\Sigma_\mathrm{heat}^\mathrm{N,sc}$ and $\Sigma_\mathrm{heat}^\mathrm{N,sf}$, can be written similarly as
\begin{align}
\Sigma_\mathrm{heat}^\mathrm{N,sc,P}=\int_{-d}^{0}\sigma_\mathrm{heat}^\mathrm{sc,P}dz
=\frac{1+\eta}{2}\Sigma_\mathrm{heat}^\mathrm{N,P}\label{sigmanpsc}\\
\Sigma_\mathrm{heat}^\mathrm{N,sf,P}=\int_{-d}^{0}\sigma_\mathrm{heat}^\mathrm{sf,P}dz
=\frac{1-\eta}{2}\Sigma_\mathrm{heat}^\mathrm{N,P}\label{sigmanpsf}
\end{align}
However, their relative magnitude is switched in comparison to the AP alignment and the spin-conserving scattering becomes dominant in the regime $2d/l_\mathrm{sf}^\mathrm{N}\ll{1}$ (see Fig.~\ref{fig-relative}). This behavior can be understood in a similar way to the AP configuration. The low-order expansion of spin current $J_\mathrm{spin}$ [see Eq.~\eqref{jspinp}] is independent of position. Then $\Delta\mu$ approaches zero because it is proportional to the gradient of $J_\mathrm{spin}$ according to Eq. \eqref{jspinderivative}. Therefore, $\sigma_\mathrm{heat}^\mathrm{sf,P}$ approaches zero in this regime according to Eq.~\eqref{heatsf}.

In the opposite regime, $2\xi=2d/l_\mathrm{sf}^\mathrm{N}\gg{1}$, the integral heat generation $\Sigma_\mathrm{heat}^\mathrm{N,sc}$ and $\Sigma_\mathrm{heat}^\mathrm{N,sf}$ approach the same value for both P and AP alignments as shown in Fig.~\ref{fig-relative}. This means that the spin-conserving scattering has the same contribution as the spin-flip one when the NM-layer thickness becomes much larger than the spin-diffusion length. The semi-infinite case discussed above is recovered in the limit of thick layer.

\subsection{Boundary conditions and heat generation at interfaces \label{sub:Bulk-spin-dependent-scattering-1}}

\begin{figure}
\includegraphics[width=0.48\textwidth]{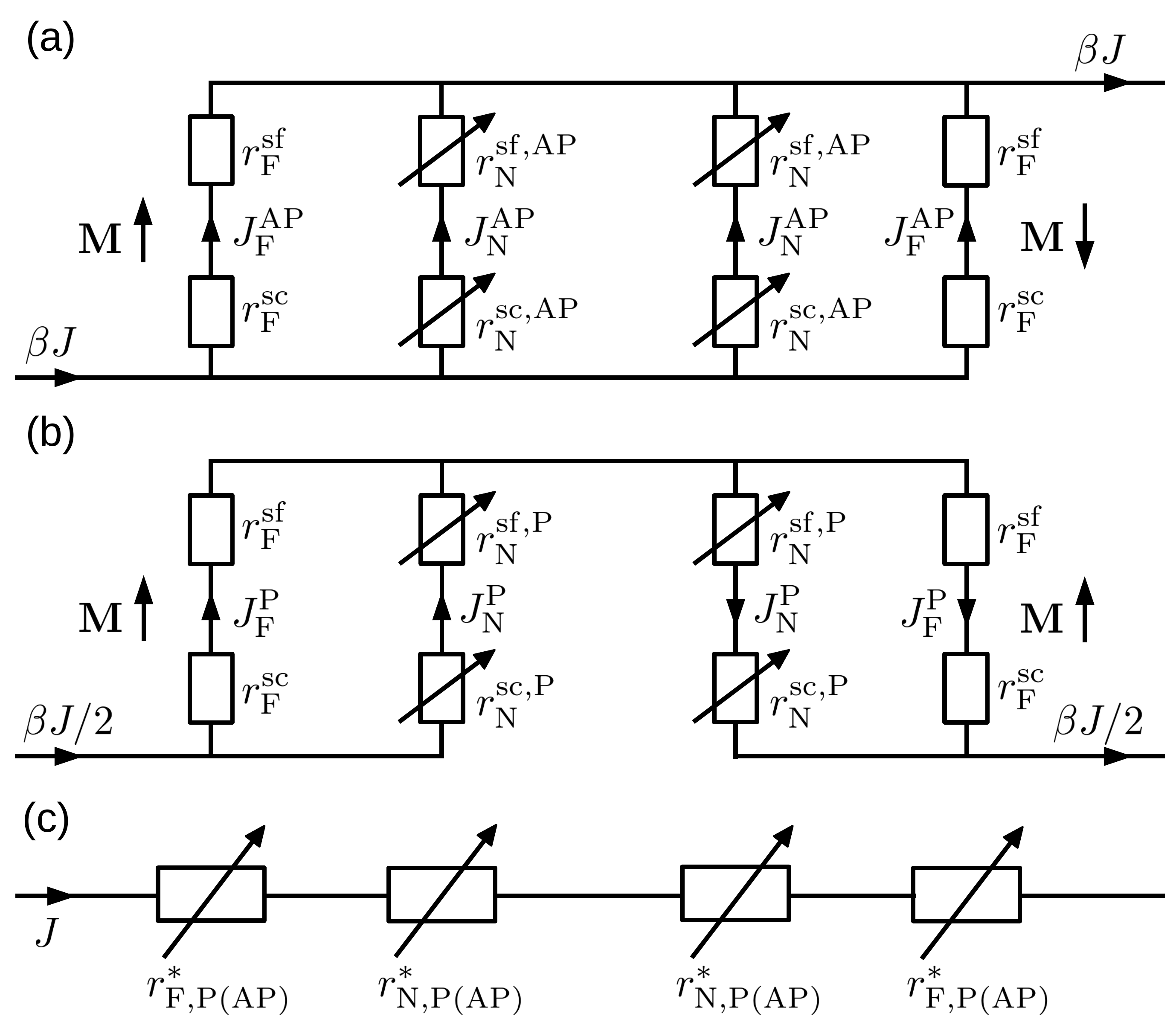}
\caption{\label{fig:circuit}Effective circuits for the spin-dependent heat generation in a spin valve with AP and P configuration neglecting the interface resistance. (a) In the AP configuration, the resistance $r_\mathrm{F}^\mathrm{sf}$ and $r_\mathrm{F}^\mathrm{sc}$ have the same value $2r_\mathrm{F}$. The variable resistances $r_\mathrm{N}^\mathrm{sc,AP}$ and $r_\mathrm{N}^\mathrm{sf,AP}$ are defined as $2r_\mathrm{N}^\mathrm{AP}(1-\eta)$ and $2r_\mathrm{N}^\mathrm{AP}(1+\eta)$, respectively. They vary with the thickness of the NM layer. (b) In the P configuration, $r_\mathrm{F}^\mathrm{sf}$ and $r_\mathrm{F}^\mathrm{sc}$ are the same as those in (a), while $r_\mathrm{N}^\mathrm{sc,P}$ and $r_\mathrm{N}^\mathrm{sf,P}$ are defined as $2r_\mathrm{N}^\mathrm{P}(1+\eta)$ and $2r_\mathrm{N}^\mathrm{P}(1-\eta)$, respectively. (c) The equivalent circuit of (a) and (b). The variable effective resistances are defined by Eqs.~\eqref{effrf} and \eqref{effrn}.}
\end{figure}

If the FM/NM interface is an Ohmic contact, the interface resistance is usually negligible in comparison to the bulk term. Using the general solutions and boundary conditions without interface resistance in ~\ref{sec:Appendix:-General-Expression}, we can determine the dimensionless parameters
\begin{align}
\alpha_\mathrm{F}^\mathrm{P(AP)}&=\frac{\beta r_\mathrm{N}^\mathrm{P(AP)}}{r_\mathrm{F}+r_\mathrm{N}^\mathrm{P(AP)}}\label{alphafnoir}\\
\alpha_\mathrm{N}^\mathrm{P(AP)}&=\frac{\beta r_\mathrm{F}}{r_\mathrm{F}+r_\mathrm{N}^\mathrm{P(AP)}}\label{alphannoir}
\end{align}
which satisfy the identity $\alpha_\mathrm{F}^\mathrm{P(AP)}+\alpha_\mathrm{N}^\mathrm{P(AP)}=\beta$. These parameters can be interpreted by the effective circuits in Figs.~\ref{fig:circuit}(a) and (b).

In the AP configuration shown by Fig.~\ref{fig:circuit}(a), the current of density $\beta{J}/2$ flows from the spin-down to the spin-up channel in either half of the spin valve. It models the electron-number current (spin flux) flowing inversely due to the spin-flip scattering in the FM and NM layers. We have the current density $J_\mathrm{F}^\mathrm{AP}=J\alpha_\mathrm{F}^\mathrm{AP}/2$ and $J_\mathrm{N}^\mathrm{AP}=J\alpha_\mathrm{N}^\mathrm{AP}/2$ by using simple circuit theorem. The heat generation due to the spin-flip scattering in the FM layer and either half of the NM layer is modeled by the Joule heating of resistances $r_\mathrm{F}^\mathrm{sf}$ and $r_\mathrm{N}^\mathrm{sf,AP}$, respectively. Similarly, the resistances $r_\mathrm{F}^\mathrm{sc}$ and $r_\mathrm{N}^\mathrm{sc,AP}$ model the heat generation due to the spin-conserving scattering. One can easily recover Eqs.~\eqref{totalh1}, \eqref{sigmanapsc}, and \eqref{sigmanapsf} by applying Joule's law to the circuit.

In the P configuration as shown by Fig.~\ref{fig:circuit}(b), the current of density $\beta{J}/2$ flows from the spin-down to the spin-up channel in the left half of the spin valve, which is similar to the AP configuration. However, the current of density $\beta{J}/2$ flows inversely in the right half of the spin valve because the sign of the spin accumulation and associated spin relaxation is switched in this half. One can find the current density $J_\mathrm{F}^\mathrm{P}=J\alpha_\mathrm{F}^\mathrm{P}/2$ and $J_\mathrm{N}^\mathrm{P}=J\alpha_\mathrm{N}^\mathrm{P}/2$ from the circuit. Then one can recover Eqs.~\eqref{totalh1}, \eqref{sigmanpsc}, and \eqref{sigmanpsf} by applying Joule's law to the circuit. Moreover, we also constructed an equivalent circuit, shown in Fig.~\ref{fig:circuit}(c), for the circuits in Figs.~\ref{fig:circuit}(a) and (b). The equivalent circuit has the same heat generation as the original one. However, the current density passing the equivalent circuit is the total current density $J$ instead so that it can be connected in series with the nominal resistances (not shown in the figure).

If the interface resistance has to be taken into account, the dimensionless parameters can be determined similarly
\begin{align}
\alpha_\mathrm{F}^\mathrm{P(AP)}&=\frac{\beta\left[r_\mathrm{N}^\mathrm{P(AP)}
+r_\mathrm{b}^\ast\right]-\gamma{r}_\mathrm{b}^\ast}
{r_\mathrm{F}+r_\mathrm{b}^\ast+r_\mathrm{N}^\mathrm{P(AP)}}\label{alphaf}\\
\alpha_\mathrm{N}^\mathrm{P(AP)}&=\frac{\beta{r}_\mathrm{F}+\gamma{r}_\mathrm{b}^\ast}
{r_\mathrm{F}+r_\mathrm{b}^\ast+r_\mathrm{N}^\mathrm{P(AP)}}\label{alphan}
\end{align}
which still satisfy the identity $\alpha_\mathrm{F}^\mathrm{P(AP)}+\alpha_\mathrm{N}^\mathrm{P(AP)}=\beta$. Equations \eqref{alphafnoir} and \eqref{alphannoir} can be recovered if $r_\mathrm{b}^\ast$ is neglected in Eqs.~\eqref{alphaf} and \eqref{alphan}.

According to Eq.~\eqref{heatcspin}, we can write the spin-dependent heat generation due to the left interface resistance as
\begin{equation}\label{heatcl}
\Sigma_\mathrm{heat}^\mathrm{C,P(AP)}(-d)=r_\mathrm{b}^\ast
\left[J_\mathrm{spin}^\mathrm{sa,P(AP)}(-d)\right]^2
\end{equation}
where the spin-up electrons are in the minority channel for both P and AP configuration, and $J_\mathrm{spin}^\mathrm{sa,P(AP)}(-d)$ is defined by Eq.~\eqref{spincurrentc}. To show the meaning of this result, we consider the current density driven by the electric field (ef) alone
\begin{equation}
J_\pm^\mathrm{ef}(-d)=\frac{1\mp\gamma}{2}J
\end{equation}
where $J$ passes through $r_+=2r_\mathrm{b}^\ast(1+\gamma)$ and $r_-=2r_\mathrm{b}^\ast(1-\gamma)$ in parallel. One can easily verify that the nominal heat generation due to the interface resistance, $[J_+^\mathrm{ef}(-d)]^2r_++[J_-^\mathrm{ef}(-d)]^2r_-$, is given by $J\Delta{V}_0$ [see Eq.~\eqref{rifnominal}]. The spin current resulting from the electric field alone is $J_\mathrm{spin}^\mathrm{ef,P(AP)}=J_+^\mathrm{ef}(-d)-J_-^\mathrm{ef}(-d)=-\gamma{J}$. The spin current due to the change of spin accumulation (sa) across the interface is the difference between $J_\mathrm{spin}^\mathrm{P(AP)}(-d)$ and $J_\mathrm{spin}^\mathrm{ef,P(AP)}(-d)$, that is, $J_\mathrm{spin}^\mathrm{sa,P(AP)}(-d)=J_\mathrm{spin}^\mathrm{P(AP)}(-d)+\gamma{J}$ already used in Eq.~\eqref{heatcl}. The spin-dependent heat generation is solely caused by the spin-conserving scattering in the presence of spin accumulation change across the interface because the spin-flip scattering is neglected.

Similarly, the spin-dependent heat generation at the right interface can be written as
\begin{equation}\label{heatcr}
\Sigma_\mathrm{heat}^\mathrm{C,P(AP)}(d)
=r_\mathrm{b}^\ast\left[J_\mathrm{spin}^\mathrm{P(AP)}(d)+(-)\gamma{J}\right]^{2}
\end{equation}
where the spin-up channel is the minority (majority) one in P (AP) alignment. Substituting $J_\mathrm{spin}^\mathrm{P(AP)}(\pm{d})$ into Eqs.~\eqref{heatcl} and \eqref{heatcr}, we can also write the spin-dependent heat generation at each interface as
\begin{equation}
\Sigma_\mathrm{heat}^\mathrm{C,P(AP)}(\pm{d})
=r_\mathrm{b}^\ast\left[\alpha_\mathrm{C}^\mathrm{P(AP)}J\right]^{2}
\label{eq:Q-interface}
\end{equation}
where
\begin{equation}\label{alphac}
\alpha_\mathrm{C}^\mathrm{P(AP)}=\gamma-\alpha_\mathrm{N}^\mathrm{P(AP)}
=\frac{\gamma\left[r_\mathrm{F}+r_\mathrm{N}^\mathrm{P(AP)}\right]
-\beta{r}_\mathrm{F}}{r_\mathrm{F}+r_\mathrm{b}^\ast+r_\mathrm{N}^\mathrm{P(AP)}}
\end{equation}
is also a dimensionless parameter.

\section{Relation to the spin-coupled interface resistance\label{sec:Relation-with-the}}

This section will give a specific discussion on the relation between the spin-dependent heat generation and the `Joule heating' of the spin-coupled interface resistance $r_\mathrm{SI}$ in the spin valves. We will show the validity and limitation of $r_\mathrm{SI}$ in describing the heat generation, and then introduce effective resistance as an alternative.

\begin{figure}
\includegraphics[width=0.48\textwidth]{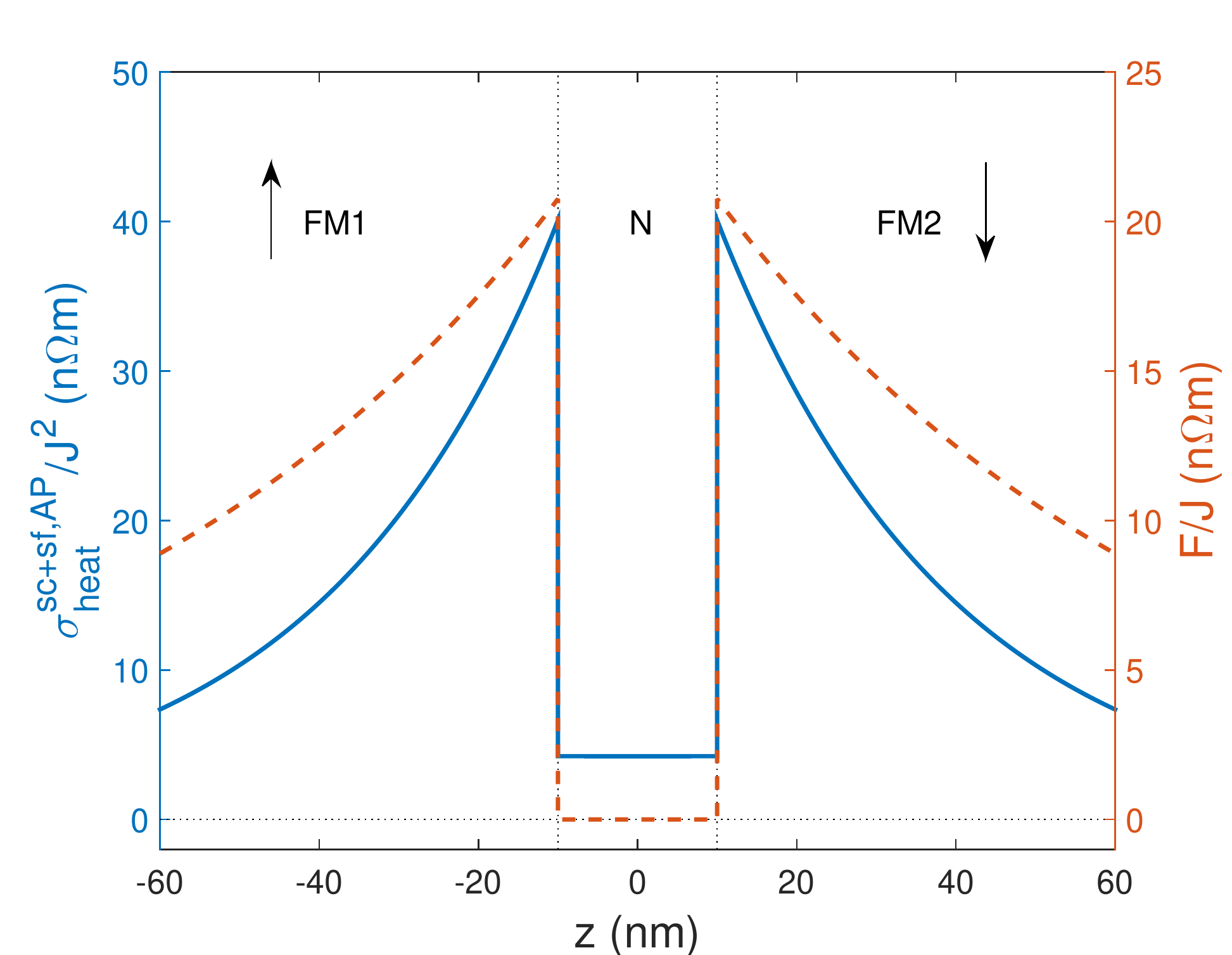}
\caption{\label{fig:heat and electric field}The spin-dependent heat generation (blue-solid curve) and the extra field (red-dashed curve) as functions of position $z$ in a spin valve with AP configuration. The two FM layers are made of Co ($\rho_\mathrm{F}^\ast=86~\mathrm{n}\Omega~\mathrm{m}$, $l_\mathrm{sf}^\mathrm{F}=59~\mathrm{nm}$, $\beta=0.5$) and the NM layer of Cu ($r_\mathrm{N}=7~\mathrm{n}\Omega~\mathrm{m}$,
$l_\mathrm{sf}^\mathrm{N}=450~\mathrm{nm}$). The thickness of the NM layer is $20$ nm and the interface resistance is neglected for simplicity.}
\end{figure}

\subsection{Validity of the spin-coupled interface resistance}

The spin-coupled interface resistances can be calculated by using Eqs.~\eqref{rsi} and \eqref{rsiif}. Substituting Eqs.~\eqref{eq:F:FM1}, \eqref{eq:F:N}, and \eqref{eq:F:FM2} into Eq.~\eqref{rsi}, we have
\begin{align}
r_\mathrm{SI}^\mathrm{F,P(AP)}&=\beta{r}_\mathrm{F}\alpha_\mathrm{F}^\mathrm{P(AP)}\label{rsif}\\
r_\mathrm{SI}^\mathrm{N,P(AP)}&=0\label{rsin}
\end{align}
where $r_\mathrm{SI}^\mathrm{F,P(AP)}$ is for either of the two FM layers and $r_\mathrm{SI}^\mathrm{N,P(AP)}$ for the NM layer. Substituting \eqref{deltamu} into Eq.~\eqref{rsiif} and using Eq. \eqref{alphac}, we can write $r_\mathrm{SI}$ due to the interface as
\begin{equation}\label{rsicl}
r_\mathrm{SI}^\mathrm{C,P(AP)}
=\gamma{r}_\mathrm{b}^\ast\alpha_\mathrm{C}^\mathrm{P(AP)}
\end{equation}
which has the same value at the two interfaces of the spin valve.

One may be tempted to interpret the spin-dependent heat generation as the Joule heating of the spin-coupled interface resistance. According to Eq.~\eqref{requirement}, this is possible only if $J_\mathrm{spin}\Delta\mu$ has the same value at the two ends of the segment under consideration. The spin valve has at least three points satisfying this requirement: $z=\pm\infty$ and $z=0$.~\cite{Fert1993} Thus Joule's law is valid (at least) in the following three segments: $-\infty<z<0$, $0<z<\infty$, and of course $-\infty<z<\infty$. To be specific, we will consider the left half ($-\infty<z<0$) of the spin valve. By using Eq.~\eqref{heatspin2}, we can write \emph{formally} the spin-dependent heat generation as the Joule heating of the corresponding $r_\mathrm{SI}$
\begin{equation}\label{eq:totalq}
\Sigma_\mathrm{heat}^\mathrm{spin,P(AP)}
=J^{2}\left[r_\mathrm{SI}^\mathrm{F,P(AP)}+r_\mathrm{SI}^\mathrm{C,P(AP)}\right]
\end{equation}
where we have used $r_\mathrm{SI}^\mathrm{N,P(AP)}=0$. Note that this result does not hold for arbitrary segment and $r_\mathrm{SI}J^2$ cannot be regarded as the heat generation either because it may be negative in certain situation. Its real meaning will be discussed in the following.

We will focus on the left half of the spin valve and consider only the P alignment without loss of generality. Substituting $J_\mathrm{spin}=-\beta{J}+J_\mathrm{spin}^\mathrm{exp}$ into Eq.~\eqref{totalhg}, we have
\begin{equation}\label{heathsv}
\begin{split}
\Sigma_\mathrm{heat}&=J\left(\Delta{V}_0^\mathrm{F}+\Delta{V}_0^\mathrm{C}
+\Delta{V}_0^\mathrm{N}\right)\\
&+J\left(\Delta{V}_\mathrm{I}^\mathrm{F}
+\Delta{V}_\mathrm{I}^\mathrm{C}\right)-\Delta{E}_\mathrm{cp}
+\Sigma_\mathrm{heat}^\mathrm{spin,P}
\end{split}
\end{equation}
where we have introduced
\begin{equation}
\Delta{E}_\mathrm{cp}=\frac{\beta{J}}{e}\Delta\mu(z_\mathrm{C}^-)
+\frac{\gamma{J}}{e}\left[\Delta\mu(z_\mathrm{C}^+)-\Delta\mu(z_\mathrm{C}^-)\right]
\end{equation}
and
\begin{equation}\label{heatspin3}
\begin{split}
\Sigma_\mathrm{heat}^\mathrm{spin,P}&=
\frac{J_\mathrm{spin}^\mathrm{exp}(z_\mathrm{C}^-)}{e}\Delta\mu(z_\mathrm{C}^-)
-\frac{J_\mathrm{spin}^\mathrm{exp}(z_\mathrm{C}^+)}{e}\Delta\mu(z_\mathrm{C}^+)\\
&+\frac{\gamma{J}+J_\mathrm{spin}(z_\mathrm{C})}{e}
\left[\Delta\mu(z_\mathrm{C}^+)-\Delta\mu(z_\mathrm{C}^-)\right]
\end{split}
\end{equation}
Using Eqs.~\eqref{bcif1}, \eqref{bcif2}, and \eqref{deltamu}, one can easily verify that Eq.~\eqref{heatspin3} is the spin-dependent heat generation defined in Eqs.~\eqref{heatspin} and \eqref{eq:totalq}. Note that $J_\mathrm{spin}^\mathrm{exp}$ is discontinuous at the interface located at $z=z_\mathrm{C}$ although $J_\mathrm{spin}$ is continuous. Comparing Eqs.~\eqref{heattotal2} and \eqref{heathsv}, one has
\begin{equation}
J\left(\Delta{V}_\mathrm{I}^\mathrm{F}
+\Delta{V}_\mathrm{I}^\mathrm{C}\right)=\Delta{E}_\mathrm{cp}
\end{equation}
which is valid in \emph{arbitrary} segment of the spin valve. More specifically, we also have
\begin{align}
J\Delta{V}_\mathrm{I}^\mathrm{F}&=\frac{\beta{J}}{e}\Delta\mu(z_\mathrm{C}^-)\label{deltavif}\\
J\Delta{V}_\mathrm{I}^\mathrm{C}&=\frac{\gamma{J}}{e}
\left[\Delta\mu(z_\mathrm{C}^+)-\Delta\mu(z_\mathrm{C}^-)\right]\label{deltavic}
\end{align}
where Eqs.~\eqref{rsi} and \eqref{rsiif} have been used. On the other hand, using Eq.~\eqref{eq:totalq} and $J\left(\Delta{V}_\mathrm{I}^\mathrm{F}
+\Delta{V}_\mathrm{I}^\mathrm{C}\right)=J^{2}\left[r_\mathrm{SI}^\mathrm{F,P}
+r_\mathrm{SI}^\mathrm{C,P}\right]$, we also have
\begin{equation}
\Sigma_\mathrm{heat}^\mathrm{spin,P}=\Delta{E}_\mathrm{cp}
\end{equation}
However, this relation is valid only if $J_\mathrm{spin}\Delta\mu$ has the same value at the two terminals of the segment. Therefore, $J(\Delta{V}_\mathrm{I}^\mathrm{F}+\Delta{V}_\mathrm{I}^\mathrm{C})$ is more closely related to $\Delta{E}_\mathrm{cp}$ than $\Sigma_\mathrm{heat}^\mathrm{spin,P}$. It is crucial to figure out the meaning of $J(\Delta{V}_\mathrm{I}^\mathrm{F}+\Delta{V}_\mathrm{I}^\mathrm{C})$ and $\Delta{E}_\mathrm{cp}$.

According to the definition of $\Delta{V}_\mathrm{I}^\mathrm{F}$ [see Eq.~\eqref{rsi}], $J\Delta{V}_\mathrm{I}^\mathrm{F}$ should be regarded as the work done by the extra field, which is dominated by the electrostatic field.~\cite{Zhu14} We stress that this work may be negative when the interface resistance is included. To show this feature, we rewrite $J\Delta{V}_\mathrm{I}^\mathrm{F}$ as
\begin{equation}
J\Delta{V}_\mathrm{I}^\mathrm{F}=\beta{r}_\mathrm{F}\alpha_\mathrm{F}^\mathrm{P}J^2
\end{equation}
where $\alpha_\mathrm{F}^\mathrm{P}$ is given by Eq.~\eqref{alphaf}. By choosing the various parameters properly, one can make $\alpha_\mathrm{F}^\mathrm{P}$ negative. Then $r_\mathrm{SI}^\mathrm{F,P}=\Delta{V}_\mathrm{I}^\mathrm{F}/J$ also becomes negative and so it is ill-defined. Similarly, $J\Delta{V}_\mathrm{I}^\mathrm{C}$ stands for the work done by the extra field across the interface and it may also be negative. Using Eq.~\eqref{alphac}, we have
\begin{equation}
J\Delta{V}_\mathrm{I}^\mathrm{C}=\gamma{r}_\mathrm{b}^\ast\alpha_\mathrm{C}^\mathrm{P}J^2
\end{equation}
where $\alpha_\mathrm{C}^\mathrm{P}$ may be negative if the various parameters are chosen properly. Then $r_\mathrm{SI}^\mathrm{C,P}=\Delta{V}_\mathrm{I}^\mathrm{C}/J$ also becomes negative. Therefore, $J(\Delta{V}_\mathrm{I}^\mathrm{F}+\Delta{V}_\mathrm{I}^\mathrm{C})$ should be regarded as the work done by the extra field instead of the Joule heating of the spin-coupled interface resistance although they are equal in some special segments, for example, $\Sigma_\mathrm{heat}^\mathrm{spin,P}=J^2r_\mathrm{SI}^\mathrm{F,P}+J^2r_\mathrm{SI}^\mathrm{C,P}$ in the left half of the spin valve.

To show the meaning of $\Delta{E}_\mathrm{cp}$, we rewrite its first term as
\begin{equation}\label{2ndterm}
\frac{\beta{J}}{e}\Delta\mu(z_\mathrm{C}^-)
=\frac{J_{+}^\mathrm{bulk}}{-e}\left[\mu_{+}(z_\mathrm{C}^-)-\mu_0\right]
+\frac{J_{-}^\mathrm{bulk}}{-e}\left[\mu_{-}(z_\mathrm{C}^-)-\mu_0\right]
\end{equation}
where $\mu_0$ is the equilibrium chemical potential and $\mu_\pm$ the chemical potential for spin $s=\pm$, respectively. We have used the quasi-neutrality approximation, $\mu_{+}(z_\mathrm{C}^-)+\mu_{-}(z_\mathrm{C}^-)-2\mu_0=0$, \cite{Zhu14} when deriving Eq.~\eqref{2ndterm}. The two terms on the right-hand side of Eq.~\eqref{2ndterm} stand for the change of energy stored in the chemical-potential imbalance per unit time in the two spin channels, respectively, when the bulk electron-number current flows from $z_\mathrm{C}^-$ to $-\infty$. The reasonable source of the net energy change is the work done by the extra field in the FM layer according to Eq.~\eqref{deltavif}. Similarly, the second term of $\Delta{E}_\mathrm{cp}$ can be written as
\begin{equation}\label{ircpchange}
\frac{\gamma{J}}{e}\left[\Delta\mu(z_\mathrm{C}^+)-\Delta\mu(z_\mathrm{C}^-)\right]
=\frac{J_{+}^\mathrm{ef}(z_\mathrm{C})}{-e}\delta\mu_+
+\frac{J_{-}^\mathrm{ef}(z_\mathrm{C})}{-e}\delta\mu_-
\end{equation}
where we have introduced $\delta\mu_\pm=\mu_\pm(z_\mathrm{C}^+)-\mu_\pm(z_\mathrm{C}^-)$ and used the quasi-neutrality approximation. The two terms on the right-hand side of Eq.~\eqref{ircpchange} can be regarded as the change of energy stored in chemical-potential imbalance of the two spin channels, respectively, when the current driven by the electric field alone traverses the interface. This energy change is supplied by the extra field at the interface according to Eq.~\eqref{deltavic}.

The physical process can be summarized as follows. The spin-dependent heat generation due to the spin-conserving and spin-flip scattering leads to the dissipation of energy stored in the chemical-potential splitting in every layer of the spin valve including the FM layers, the interface, and the NM layer. Then this change of the chemical-potential energy is compensated by the work of the extra electric field in the FM layer and at the interface. However, the compensation process does not happen in the NM layer since there is no extra field in this layer.

\subsection{Effective resistance}

The spin-dependent heat generation in each individual layer cannot be interpreted as Joule heating of $r_\mathrm{SI}$ in this layer. This is easy to see if we consider the NM layer. It has a spin-dependent heat generation but no contribution to $r_\mathrm{SI}$ according to Eq.~(\ref{rsin}). Similar analysis shows that this result is also true in other layers and at interfaces. Moreover, $r_\mathrm{SI}^\mathrm{F,P(AP)}$ and $r_\mathrm{SI}^\mathrm{C,P(AP)}$ in Eqs.~\eqref{rsif} and \eqref{rsicl} can be negative when $\alpha_\mathrm{F}^\mathrm{P(AP)}$ and $\alpha_\mathrm{C}^\mathrm{P(AP)}$ are negative according to Eqs.~\eqref{alphaf} and \eqref{alphac}. Therefore, it is necessary to find an alternative to $r_\mathrm{SI}$ if one hopes to describe the heat generation of a single layer with Joule's law. Using Eqs.~\eqref{totalh1}, \eqref{totalhnmap}, \eqref{totalhnmp}, and \eqref{eq:Q-interface}, we can write the spin-dependent part of the integral heat generation in each layer as
\begin{align}
\Sigma_\mathrm{heat}^\mathrm{F,P(AP)}&=J^2r_\mathrm{F,P(AP)}^\ast\label{eq:Q-F-expf}\\
\Sigma_\mathrm{heat}^\mathrm{N,P(AP)}&=J^2r_\mathrm{N,P(AP)}^\ast\label{eq:Q-F-expn}\\
\Sigma_\mathrm{heat}^\mathrm{C,P(AP)}&=J^2r_\mathrm{C,P(AP)}^\ast\label{eq:Q-F-expc}
\end{align}
where we have introduced the effective resistances
\begin{align}
r_\mathrm{F,P(AP)}^\ast&=r_\mathrm{F}\left[\alpha_\mathrm{F}^\mathrm{P(AP)}\right]^{2}\label{effrf}\\
r_\mathrm{N,P(AP)}^\ast&=r_\mathrm{N}^\mathrm{P(AP)}
\left[\alpha_\mathrm{N}^\mathrm{P(AP)}\right]^{2}\label{effrn}\\
r_\mathrm{C,P(AP)}^\ast&=r_\mathrm{b}^\ast\left[\alpha_\mathrm{C}^\mathrm{P(AP)}\right]^{2}\label{effrc}
\end{align}
for one FM layer, half of the NM layer, and one FM/NM interface, respectively. The effective resistances $r_\mathrm{F,P(AP)}^\ast$ and $r_\mathrm{N,P(AP)}^\ast$ can be understood by constructing an equivalent circuit as shown in Fig.~\ref{fig:circuit}(c). Comparing Eqs.~\eqref{eq:totalq} with the sum of Eqs.~\eqref{eq:Q-F-expf}, \eqref{eq:Q-F-expn}, and \eqref{eq:Q-F-expc}, we have
\begin{equation}
r_\mathrm{SI}^\mathrm{P(AP)}=r_\mathrm{F,P(AP)}^\ast+r_\mathrm{N,P(AP)}^\ast+r_\mathrm{C,P(AP)}^\ast
\end{equation}
where we also have $r_\mathrm{SI}^\mathrm{P(AP)}=r_\mathrm{SI}^\mathrm{F,P(AP)}+r_\mathrm{SI}^\mathrm{C,P(AP)}$. Two resistances have been introduced for each layer: the effective resistance and the spin-coupled interface resistance. In general, they are not equal to each other in a single layer or at an interface. The most obvious example is the NM layer, where $r_\mathrm{N,P(AP)}^\ast$ has a finite value whenever $d\neq{0}$ while $r_\mathrm{SI}^\mathrm{N}$ is always zero.

\begin{figure}
\includegraphics[width=0.48\textwidth]{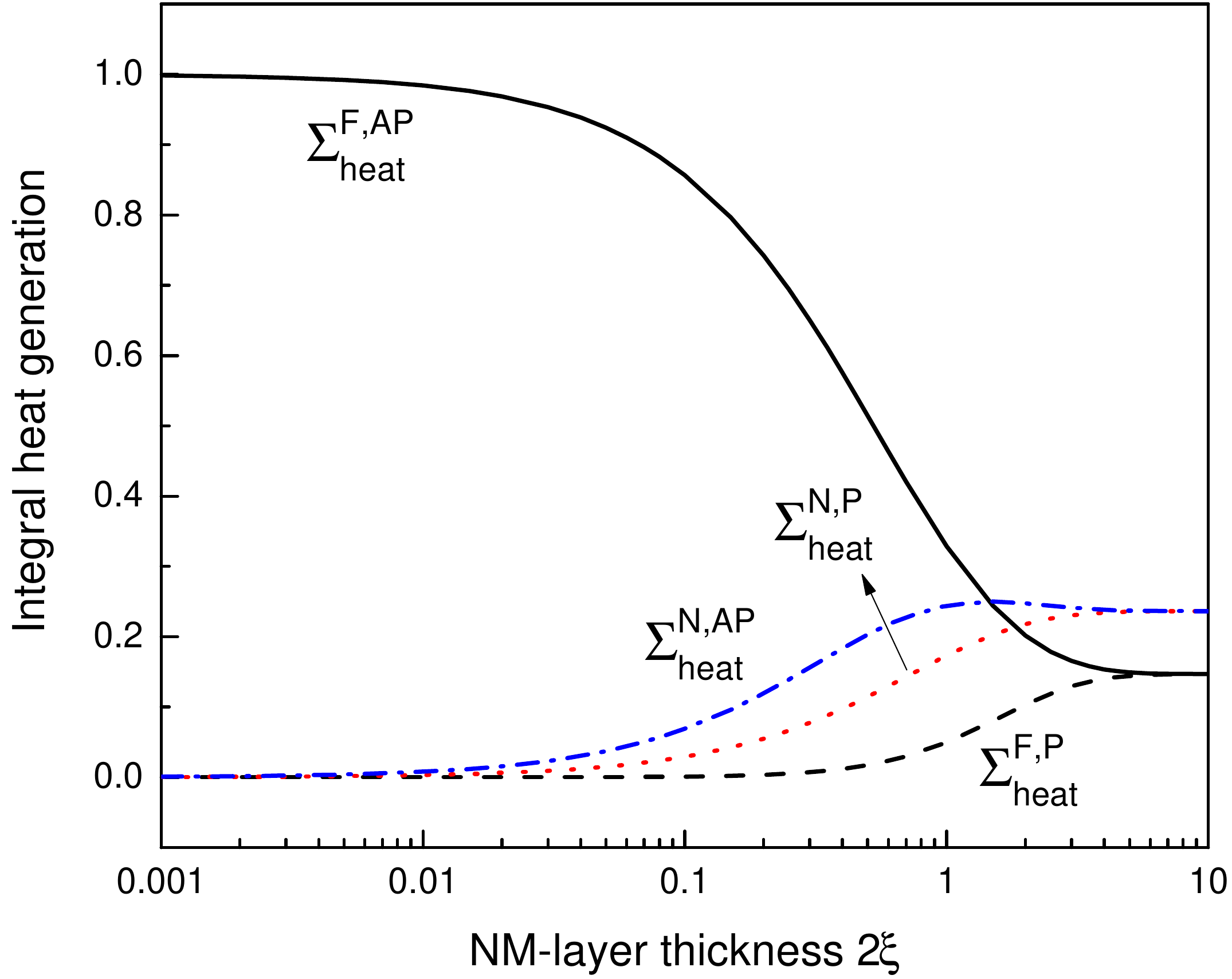}
\caption{\label{fig-relative2}The spin-dependent part of the integral spin-dependent heat generation (in unit of $\Sigma_\mathrm{heat}^\mathrm{F,AP}$) in the FM and NM layers as functions of the NM-layer thickness (in unit of $l_\mathrm{sf}^\mathrm{N}$). We used the same parameters as those of Fig.~\ref{fig:heat and electric field}. }
\end{figure}

\begin{figure}
\includegraphics[width=0.48\textwidth]{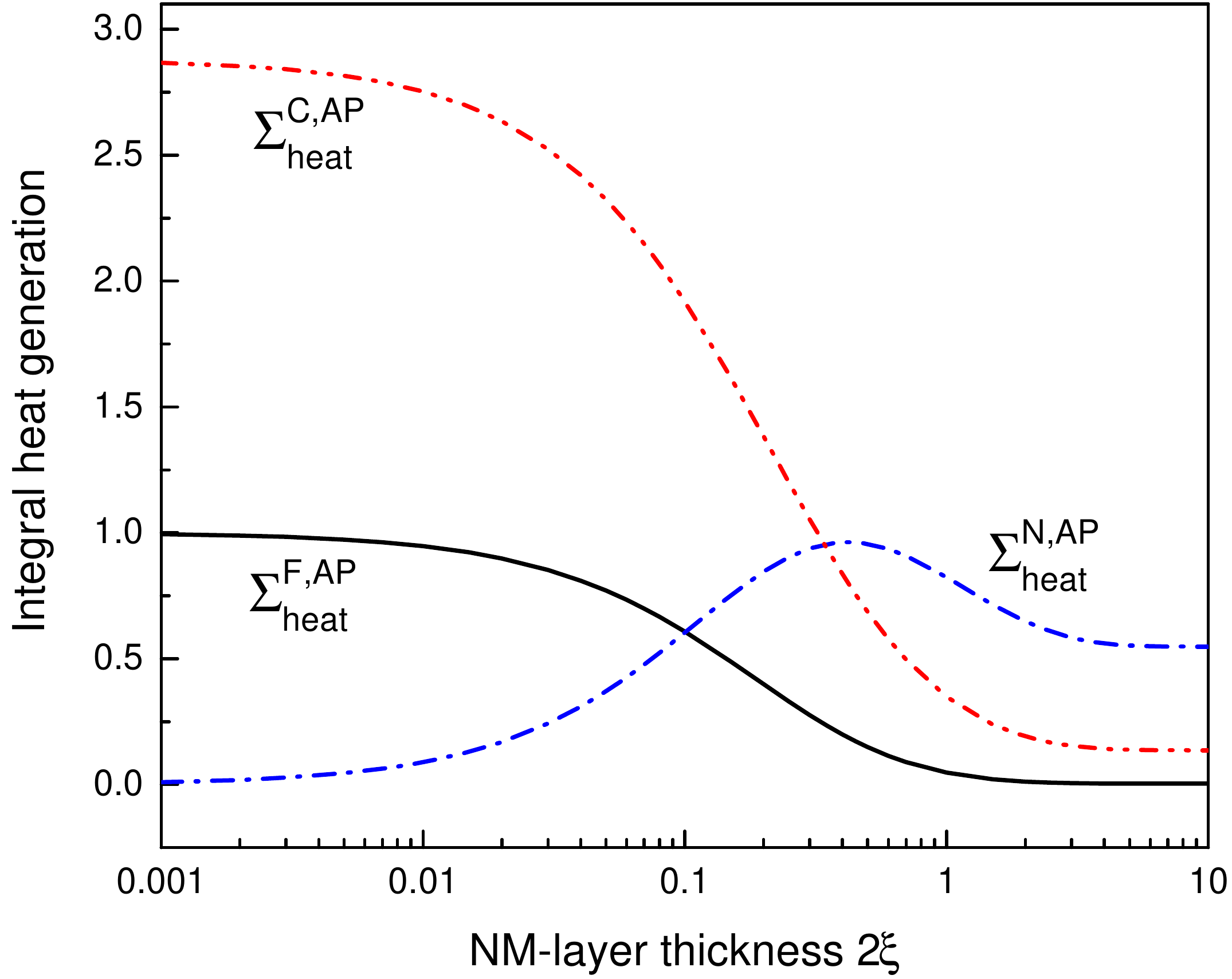}
\caption{\label{fig-relative3}The variation of the integral spin-dependent heat generation (in unit of $\Sigma_\mathrm{heat}^\mathrm{F,AP}$) with the NM-layer thickness (in unit of $l_\mathrm{sf}^\mathrm{N}$) for the AP alignment. The solid and dash-dot lines are for the FM1 layer and the left half of the NM layer as in Fig.~\ref{fig-relative2}. The contribution of the interface resistance ($r_\mathrm{b}^\ast=2r_\mathrm{F}$ and $\gamma=0.6$) is shown by the dash-dot-dot line.}
\end{figure}

Figure \ref{fig-relative2} shows the quantitative results for the integral heat generation without interface resistance. The curves can also be regarded as the variation of the effective resistances defined in Eqs.~\eqref{effrf} and \eqref{effrn} since they are proportional to the integral heat generation. In the regime $2\xi\ll1$, $\Sigma_\mathrm{heat}^\mathrm{F,AP}$ approaches a positive constant $J^2\beta^2r_\mathrm{F}$, while $\Sigma_\mathrm{heat}^\mathrm{F,P}$, $\Sigma_\mathrm{heat}^\mathrm{N,AP}$, and $\Sigma_\mathrm{heat}^\mathrm{N,P}$ approach zero. This behavior indicates the existence of the magneto-heating effect: different heat generation in P and AP configurations. In the limit of $2\xi\rightarrow\infty$, the difference between P and AP alignments disappears. The limits of the heat generation in the FM and NM layers depend on $r_\mathrm{F}$ and $r_\mathrm{N}$. If $r_\mathrm{F}$ is equal to $r_\mathrm{N}$, all the curves approach the same limit. Moreover, under the condition $r_\mathrm{F}=r_\mathrm{N}$, we also have $\Sigma_\mathrm{heat}^\mathrm{N,AP}=\Sigma_\mathrm{heat}^\mathrm{N,P}$ for arbitrary NM-layer thickness according to Eqs.~\eqref{totalhnmap} and \eqref{totalhnmp}.

When the interface resistance is taken into account, the spin-dependent part of the integral heat generation exhibits several important features as shown in Fig.~\ref{fig-relative3}. The contribution of the FM layer may decrease to zero if the NM-layer thickness increases to a certain value. Then we have $\alpha_\mathrm{F}^\mathrm{AP}=0$ according to Eq.~\eqref{totalh1} and the spin accumulation disappears together with the additional field in the FM layer. In this case, the loss of potential energy due to heat generation can only be compensated by the additional field at the interface. Moreover, the contribution of the NM layer may exceed that from the FM layer even when $2\xi$ is small enough that the magneto-heating effect is still remarkable. This indicates the possibility of allocating heat generation in different layers by engineering the interface resistance.

\section{Conclusions\label{sec:Conclusions-1}}

Our analytical results show that the spin-dependent heat generation is due to two mechanisms: the spin-flip and spin-conserving scattering. In the presence of spin accumulation, heat is generated when electrons undergo transitions between the two spin channels with different chemical potential via the spin-flip scattering. On the other hand, with the existence of spin accumulation gradient, spin-conserving scattering can also lead to heat generation when electrons move to positions with lower chemical potential in the same spin channel. The two mechanisms have equal contributions in semi-infinite layers. However, in the NM layer of a thickness much shorter than its spin-diffusion length, the spin-dependent heat generation is dominated by the spin-flip scattering in the AP configuration, and by the spin-conserving scattering in the P configuration.

We proved in general that the spin-dependent heat generation is equal to the `Joule heating' of the spin-coupled interface resistance ($r_\mathrm{SI}$) only in some special segment. The concept of $r_\mathrm{SI}$ has another limitation: it may be negative in some cases when it is defined in an individual layer. Therefore, $J^2r_\mathrm{SI}$ should be interpreted as the work done by the extra field in the FM layers and at interfaces instead of Joule heating. It converts into the energy stored in the chemical-potential splitting, which in turn compensates the spin-dependent energy dissipation in all the layers including the NM layer. Effective resistances and associated circuits are also introduced to overcome the limitation of $r_\mathrm{SI}$ in describing heat generation.

\section{Acknowledgments}

We thank Prof.~Suzuki and Prof.~Tulapurkar for fruitful discussions. This work was supported by National Natural Science Foundation of China [grant numbers~11404013, 11605003, 61405003, 11174020, 11474012]; Beijing Natural Science Foundation [grant number~1112007]; and 2016 Graduate Research Program of BTBU.

\appendix

\section{General Expression\label{sec:Appendix:-General-Expression}}

The current density $J_s$ and electrochemical potential $\bar\mu_s=\mu_s-eV$ of the two spin channels, $s=\pm$, satisfy the following equations
\begin{align}
\frac{e}{\sigma_{s}}\frac{\partial J_{s}}{\partial z} & =\frac{\bar{\mu}_{s}-\bar{\mu}_{-s}}{l_{s}^{2}}\label{eq:conti:spin-flip}\\
J_{s} & =\frac{\sigma_{s}}{e}\frac{\partial\bar{\mu}_{s}}{\partial z}\label{eq:conti:Ohm's law}
\end{align}
where $\sigma_s$ and $l_s$ denote the conductivity and spin-diffusion length for spin $s$, respectively.~\cite{Fert1993} These two equations can be transformed into
\begin{align}
&\frac{e}{\sigma_\pm}\frac{\partial{J}_\pm}{\partial{z}} =\pm2\frac{\Delta\mu}{l_\pm^2}\label{eq:conti:spin-flip2}\\
&J_\pm=\sigma_\pm\left(F\pm\frac{1}{e}
\frac{\partial\Delta\mu}{\partial{z}}\right)\label{eq:conti:Ohm's law2}
\end{align}
where $\Delta\mu=(\bar\mu_+-\bar\mu_-)/2$ and $F=(1/e)\partial\bar\mu/\partial{z}$. Solving these equations for the spin valve considered in Sec.~\ref{sec:Joule-heating-in-1}, we can write $\bar{\mu}_{\pm}\left(z\right)$, $J_{\pm}\left(z\right)$ and $F\left(z\right)$ in terms of $\Delta\mu$. For the AP configuration, the magnetization direction is ``up'' in the left FM layer (FM1) and ``down'' in the right FM layer (FM2). In the FM1 layer $\left(z<-d\right)$, we have
\begin{align}
\Delta\mu(z)&=er_\mathrm{F}\alpha_\mathrm{F}^\mathrm{AP}J
\exp\left[(z+d)/l_\mathrm{sf}^\mathrm{F}\right]\label{deltamufm1}\\
\bar{\mu}_{\pm}\left(z\right)&=eE_0^\mathrm{F}z
+K_{1}^\mathrm{A}\pm\left(1\pm\beta\right)\Delta\mu\\
J_{\pm}\left(z\right)&=\left(1\mp\beta\right)\frac{J}{2}\pm\frac{\Delta\mu}{2er_\mathrm{F}}\\
F\left(z\right)&=E_0^\mathrm{F}+\frac{\beta\Delta\mu}{el_\mathrm{sf}^\mathrm{F}}\label{eq:F:FM1}
\end{align}
In the NM layer $\left(-d<z<d\right)$, we have
\begin{align}
\Delta\mu(z)&=er_\mathrm{N}^\mathrm{AP}\alpha_\mathrm{N}^\mathrm{AP}J
\cosh(z/l_\mathrm{sf}^\mathrm{N})/\cosh\xi\label{deltamunm}\\
\bar{\mu}_{\pm}\left(z\right)&=eE_0^\mathrm{N}z+K_{1}^\mathrm{B}\pm\Delta\mu\\
J_{\pm}\left(z\right)&=\frac{J}{2}\pm\frac{1}{2e\rho_\mathrm{N}^\ast}
\frac{\partial\Delta\mu}{\partial{z}}\\
F\left(z\right)&=E_0^\mathrm{N}\label{eq:F:N}
\end{align}
where $E_0^\mathrm{N}=\rho_\mathrm{N}^\ast{J}$. In the FM2 layer $\left(z>d\right)$, we have
\begin{align}
\Delta\mu(z)&=er_\mathrm{F}\alpha_\mathrm{F}^\mathrm{AP}J
\exp\left[-(z-d)/l_\mathrm{sf}^\mathrm{F}\right]\label{deltamufm2}\\
\bar{\mu}_{\pm}\left(z\right)&=eE_0^\mathrm{F}z
+K_{1}^{C}\pm\left(1\mp\beta\right)\Delta\mu\\
J_{\pm}\left(z\right)&=\left(1\pm\beta\right)\frac{J}{2}\mp\frac{\Delta\mu}{2er_\mathrm{F}}\\
F\left(z\right)&=E_0^\mathrm{F}+\frac{\beta\Delta\mu}{el_\mathrm{sf}^\mathrm{F}}\label{eq:F:FM2}
\end{align}
Using these equations, one can easily derive
\begin{equation}
J_\mathrm{spin}=\mp\beta{J}\pm\frac{\Delta\mu}{er_\mathrm{F}}
\end{equation}
for the FM1 (FM2) layer, and
\begin{equation}
J_\mathrm{spin}=\frac{1}{e\rho_\mathrm{N}^\ast}\frac{\partial\Delta\mu}{\partial{z}}
=\alpha_\mathrm{N}^\mathrm{AP}J\sinh(z/l_\mathrm{sf}^\mathrm{N})/\sinh\xi
\end{equation}
for the NM layer.

In the P configuration, both FM1 and FM2 layers have ``up'' magnetization. The expressions of the various quantities can be derived similarly and thus we only list some results that will be referred to in the previous sections. In the NM layer, we have
\begin{align}
\Delta\mu(z)&=-er_\mathrm{N}^\mathrm{P}\alpha_\mathrm{N}^\mathrm{P}J
\sinh(z/l_\mathrm{sf}^\mathrm{N})/\sinh\xi\\
J_\mathrm{spin}&=-\alpha_\mathrm{N}^\mathrm{P}J
\cosh(z/l_\mathrm{sf}^\mathrm{N})/\cosh\xi\label{jspinp}
\end{align}
Note that $J_\mathrm{spin}$ has only exponential part in the NM layer. The dimensionless parameters $\alpha_\mathrm{F}^\mathrm{P(AP)}$ and $\alpha_\mathrm{N}^\mathrm{P(AP)}$ are given in Sec.~\ref{sub:Bulk-spin-dependent-scattering-1}.

The current density and electrochemical potential satisfy the boundary conditions at an interface located at $z=z_\mathrm{C}$
\begin{align}
&J_{s}\left(z_\mathrm{C}^{+}\right)-J_{s}\left(z_\mathrm{C}^{-}\right)=0\label{bcif1}\\
&\delta\bar\mu_s(z_\mathrm{C})=\bar{\mu}_{s}\left(z_\mathrm{C}^{+}\right)
-\bar{\mu}_{s}\left(z_\mathrm{C}^{-}\right)
=er_{s}J_{s}\left(z_\mathrm{C}\right)\label{bcif2}
\end{align}
where the spin-dependent interface resistance $r_{s}$ can be written as
\begin{equation}\label{bcr}
r_{\uparrow\left(\downarrow\right)}=2r_{b}^{*}\left[1-\left(+\right)\gamma\right]
\end{equation}
Here, $\gamma$ is the interfacial asymmetry coefficient and $\uparrow$ ($\downarrow$) denotes the majority (minority) spin channel. Subtracting the `$\pm$' components of Eq.~\eqref{bcif2}, we have
\begin{equation}\label{deltamu}
\Delta\mu(z_\mathrm{C}^+)-\Delta\mu(z_\mathrm{C}^-)
=eJ_\mathrm{spin}^\mathrm{ac}(z_\mathrm{C})r_\mathrm{b}^\ast
=e\left[J_\mathrm{spin}(z_\mathrm{C})\pm\gamma{J}\right]r_\mathrm{b}^\ast
\end{equation}
where we have used Eq.~\eqref{spincurrentc}. The sign `+' (`$-$') corresponds to the configuration in which the spin-up channel is the minority (majority) one. On the other hand, summing the `$\pm$' components of Eq.~\eqref{bcif2}, we have
\begin{equation}\label{deltamubar}
\bar\mu(z_\mathrm{C}^+)-\bar\mu(z_\mathrm{C}^-)
=eJ(1-\gamma^2)r_\mathrm{b}^\ast\pm
e\gamma\left[J_\mathrm{spin}(z_\mathrm{C})\pm\gamma{J}\right]r_\mathrm{b}^\ast
\end{equation}

\section{Heat generation due to spin-conserving scattering \label{spindiffusion}}

The first two terms of Eq.~\eqref{eq:entropy2} can also be rewritten in the following way. The current density $J_s$ and the electro-chemical potential $\bar\mu_s$ can be separated into bulk and exponential terms
\begin{align}
J_{\pm}&=J_{\pm}^\mathrm{bulk}+J_{\pm}^\mathrm{exp}\\
\frac{\partial\bar{\mu}_{\pm}}{\partial{z}}&
=\frac{\partial\bar{\mu}_{\pm}^\mathrm{bulk}}{\partial{z}}
+\frac{\partial\bar{\mu}_{\pm}^\mathrm{exp}}{\partial{z}}
\end{align}
Moreover, the bulk and exponential components satisfy the following relations
\begin{align}
&J_{+}^\mathrm{exp}+J_{-}^\mathrm{exp}=0\\
&\frac{\partial\bar{\mu}_{+}^\mathrm{bulk}}{\partial{z}}
=\frac{\partial\bar{\mu}_{-}^\mathrm{bulk}}{\partial{z}}
\end{align}
The sum of the first two terms in Eq. \eqref{eq:entropy2} can be written as
\begin{equation}
\frac{J_{+}}{e}\frac{\partial\bar{\mu}_{+}}{\partial z}+\frac{J_{-}}{e}\frac{\partial\bar{\mu}_{-}}{\partial z}=\frac{J^{2}}{\sigma_{+}+\sigma_{-}}+\frac{J_{+}^\mathrm{exp}}{e}
\frac{\partial\bar{\mu}_{+}^\mathrm{exp}}{\partial z}+\frac{J_{-}^\mathrm{exp}}{e}\frac{\partial\bar{\mu}_{-}^\mathrm{exp}}{\partial z}\label{eq:bulk+diff-1}
\end{equation}
Then using $\bar{\mu}_{s}\left(z\right)=\mu_{s}\left(z\right)-eV\left(z\right)$,
we have
\begin{equation}
\begin{aligned}\frac{\partial\bar{\mu}_{\pm}^{exp}}{\partial z}=\frac{\partial\mu_{\pm}^\mathrm{exp}}{\partial z}+eF^\mathrm{exp}\end{aligned}
\end{equation}
Finally, the last two terms of Eq. \eqref{eq:bulk+diff-1} can expressed
by the chemical potential
\begin{equation}
\frac{J_{+}^\mathrm{exp}}{e}\frac{\partial\bar{\mu}_{+}^\mathrm{exp}}{\partial z}+\frac{J_{-}^\mathrm{exp}}{e}\frac{\partial\bar{\mu}_{-}^\mathrm{exp}}{\partial z}=\frac{J_{+}^\mathrm{exp}}{e}\frac{\partial\mu_{+}^\mathrm{exp}}{\partial z}+\frac{J_{-}^\mathrm{exp}}{e}\frac{\partial\mu_{-}^\mathrm{exp}}{\partial z}\label{eq:diff-1}
\end{equation}
which is the two-channel form of Eq.~\eqref{heatsc}.

\section{Heat generation due to interface resistance\label{appir}}

The heat generation due to interface resistance is only caused the spin-conserving scattering and can be calculated by simply summing the two spin channels~\cite{T:Boltz2011}
\begin{equation}\label{heatif}
\Sigma_\mathrm{heat}^\mathrm{C}=
\left[J_{+}\left(z_\mathrm{C}\right)\delta\bar\mu_{+}
+J_{-}\left(z_\mathrm{C}\right)\delta\bar\mu_{-}\right]/e,
\end{equation}
where $\delta\bar\mu_s=\bar\mu_s(z_\mathrm{C}^{+})-\bar\mu_s(z_\mathrm{C}^{-})$ is the change in electrochemical potential across the (infinitesimally thin) interface. Substituting the boundary conditions, Eq.~\eqref{bcif2}, into Eq.~\eqref{heatif}, we have
\begin{equation}
\Sigma_\mathrm{heat}^\mathrm{C}
=J_{+}^{2}\left(z_\mathrm{C}\right)r_{+}+J_{-}^{2}\left(z_\mathrm{C}\right)r_{-}
\end{equation}
which is just Joule's law in the two-channel manner. It is more meaningful to write it in terms of $J$ and $J_\mathrm{spin}$
\begin{equation}
\Sigma_\mathrm{heat}^\mathrm{C}=J\Delta{V}_0^\mathrm{C}
+\Sigma_\mathrm{heat}^\mathrm{spin,C}
\end{equation}
where $\Delta{V}_0^\mathrm{C}$ and $\Sigma_\mathrm{heat}^\mathrm{spin,C}$ are defined in Eqs.~\eqref{rifnominal} and \eqref{heatcspin}, respectively.

\section*{References}

\bibliography{heatbibfile}

\end{document}